# Tunable narrowband THz generation in the organic crystal BNA

D. Pavicevic[1], M. Nishida[1], J. Song[1], M. Buzzi[1,*], A. Cavalleri[1,2,*]

[1] *Max Planck Institute for the Structure and Dynamics of Matter, Hamburg, Germany*
[2] *Department of Physics, Clarendon Laboratory, University of Oxford, Oxford OX1 3PU, United Kingdom*
e-mail: michele.buzzi@mpsd.mpg.de, andrea.cavalleri@mpsd.mpg.de

**The generation of tunable narrowband pulses is increasingly being pursued in terahertz science, for example to study the nonlinear response of individual modes of solids and molecules. Here, we extend the chirp-and-delay method to achieve collinear phase-matched difference-frequency generation in the organic crystal N-benzyl-2-methyl-4-nitroaniline (BNA-S), which results in tunable narrowband terahertz pulses. In this configuration, the fundamental frequency of a Ti:sapphire amplifier is used — eliminating the need for optical parametric amplifiers typically required for THz generation in other organic crystals. Chirped-pulse excitation suppresses multiphoton absorption in BNA, improving stability and extending crystal lifetime. The source delivers THz transients tunable from 0.25 THz to 2.1 THz with adjustable spectral width.**

Terahertz (THz) radiation in the 0.1–10 THz range provides direct access to low-energy collective excitations [1], such as spin dynamics in magnetic materials [2,3,4], plasma modes in high-$T_c$ superconductors [5,6] and low-lying vibrational excitations [7]. Conventional broadband THz sources based on optical rectification in inorganic crystals like ZnTe and GaP allow for simple collinear phase matching at pump wavelengths between 800 nm and 1500 nm, but suffer from low conversion efficiency due to their weak second-order nonlinearities [8-10]. Other nonlinear crystals like LiNbO₃ offer high nonlinearity and damage threshold, but require complex tilted-pulse-front schemes to overcome group-phase velocity mismatch between the optical pump and the generated THz pulses [11-14]. Over the past decade, organic crystals like DAST, DSTMS, BNA, and OH1 have emerged as powerful alternatives to generate intense THz pulses, combining large nonlinear coefficients and the possibility of achieving straightforward collinear phase matching [15-22].

However, the broadband THz pulses generated by optical rectification in these crystals may simultaneously drive more than one mode at a time as well as non-resonant electronic backgrounds in the material under study, obscuring mode-specific dynamics. Narrowband THz pulses with spectral widths matched to the material's excitation linewidth, would intrinsically circumvent this issue. Difference-frequency generation (DFG) between two time-delayed, chirped near infrared pulses offers a flexible route to tunable, narrowband THz generation. Initially demonstrated in ZnTe [23], this approach has since been applied to LiNbO₃ [24] and extended to organic crystals such as DSTMS, OH1 and HMQ-TMS [25-27].

In this work, we present a chirped-pulse DFG scheme for narrowband THz generation in the organic nonlinear crystal BNA. Two replicas of a chirped 785 nm pulse are recombined in an interferometer with variable delay, producing a temporally modulated intensity profile that drives inter-pulse DFG generating narrowband THz radiation. This approach eliminates the need for an optical parametric amplifier and operates directly using the fundamental output of a Ti:Sapphire laser, offering simultaneously continuous center frequency tunability from 0.25 to 2.1 THz and sub-THz bandwidths using a simple, fully collinear geometry.

Figure 1 summarizes the principles of narrowband THz pulse generation by optical rectification (Fig. 1(a-b)) and inter-pulse chirped-DFG (Fig. 1(c-d)). Figure 1a shows the spectrogram of a fully compressed driving pulse. In optical rectification all the frequency components of the optical pulse interact with each other, yielding a broadband THz pulse as a result (Fig. 1b). Fig. 1c shows the spectrograms of the two time-delayed chirped optical pulses that participate in an inter-pulse chirped-DFG process. Here, the bandwidth is determined by the closest/farthest interacting frequency components. Due to the linear chirp and their relative delay, only a subset of their frequency components can interact, yielding a narrower bandwidth THz pulse.

For a linearly chirped Gaussian pulse, the electric field can be expressed as $E(t) = E_0(t)e^{i(\omega_0 t + Ct^2)}$, where $E_0(t)$ is the pulse envelope, $\omega_0$ is the carrier frequency and $C$ is the linear chirp rate. The intensity resulting from the superposition of two replicas ($E_1(t)$ and $E_2(t)$) with delayed by a time delay $\tau$ can be written as:

$$I_{\text{opt}}(t, C, \tau) = |E_1(t) + E_2(t+\tau)|^2 = I_1(t) + I_2(t+\tau) + 2\sqrt{I_1(t)I_2(t+\tau)}\cos(\omega_0\tau + C\tau^2 + 2C\tau t). \quad (1)$$

The last term results in a time modulated intensity profile, with a frequency (beat note) $\Omega = C\tau/\pi$, which is tunable both by adjusting the time delay $\tau$ or the chirp rate C.

The generated terahertz spectra $E_{\text{THz}}(\Omega)$ are governed by the second-order non-linear polarization generated in the nonlinear crystal. It can be shown that $E_{\text{THz}}(\Omega)$ can be expressed as:

$$E_{\text{THz}}(\Omega) \propto t_e(\Omega)T_i(\omega_0)H(\Omega,\omega_0)\,\hat{I}_{\text{opt}}(\Omega, C, \tau), \quad (2)$$

where $\Omega$ is the terahertz frequency, $t_e(\Omega)$ and $T_i(\omega_0)$ are the Fresnel transmission coefficients for THz and pump field at the exit and input surface, $\hat{I}_{\text{opt}}(\Omega, \omega_0)$ is the Fourier transform of $I_{opt}(t, C, \tau)$ and $H(\Omega, \omega_0)$ is the crystal response function as defined in Ref. [26] accounting for propagation effects and phase mismatch [22,30]. The beat note of $I_{opt}(t, C, \tau)$ gives rise to a finite frequency peak in $\hat{I}_{opt}(\Omega, C, \tau)$ with a width that depends on both the time delay $\tau$ and the chirp rate C.

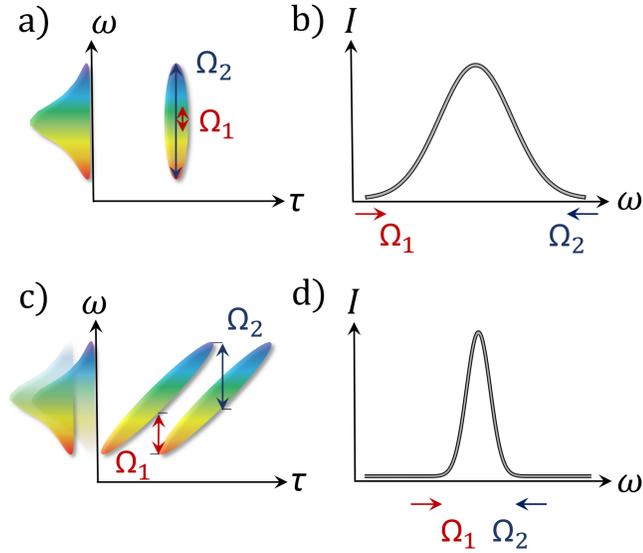

**Figure 1: (a)** Spectrogram of a Fourier transform limited pulse generating THz radiation via optical rectification. This process can be thought of as intra-pulse DFG, where THz components arise from DFG between different spectral frequencies within the pulse. $\Omega_1$ and $\Omega_2$ denote the lowest and highest THz frequencies, respectively. **(b)** Corresponding generated broadband THz spectrum. **(c)** Spectrogram of two time-delayed replicas of a chirped pulse. Here, inter-pulse DFG produces a narrowband THz spectrum as shown in **(d)**.

A schematic of the experimental setup is shown in Fig. 2. The system is based on a commercial Ti:sapphire amplifier delivering 35 fs, 785 nm pulses at 1 kHz repetition rate with ~2.5 mJ pulse energy. Chirped pulse durations of 1-4 ps were obtained by detuning the compressor grating separation away from its optimal value, corresponding to group delay dispersions (GDD) of ~35,000-110,000 fs². The duration of the chirped optical pulses was directly measured using intensity autocorrelation.

The chirped pulses entered a variable-delay Mach–Zehnder interferometer, where a first dielectric beamsplitter produced two replicas. A delay stage in one arm introduced a variable time delay $\tau$, and the beams were recombined at a second beamsplitter.

The output from one interferometer port was directed to a ~500-µm-thick BNA-S crystal (THz Innovations). After the crystal, the residual pump was blocked using a stack of high-density polyethylene and thin black polypropylene filters. The emitted THz beam was expanded with a 4x off-axis parabolic (OAP) mirror telescope and focused onto a 100 µm-thick (110)-oriented GaP crystal for electro-optic sampling (EOS). The gate pulses, derived from a split portion of the Ti:sapphire beam, were recompressed to <50 fs using an auxiliary transmission grating compressor. The delay $\tau$ was calibrated by monitoring the spectral fringes at the interferometer output. The time-delayed replicas produce fringes with spacing $\Delta\lambda \sim \lambda^2/c\tau$, independent of the chirp rate C. The inset in Figure 2 shows a representative spectrum for 4 ps chirped pulses with a delay $\tau \sim 240$ fs, corresponding to $\Delta\lambda \sim 8.5$ nm. Alignment of the focusing OAP was optimized using a commercial THz microbolometer camera, and the THz beam diameter was determined via knife-edge measurements with a pyroelectric detector. For 1 THz narrowband generation (FWHM bandwidth~0.5 THz), the THz beam diameter was ~400 µm (FWHM). With 1.2 ps long, 1.2 mJ pump pulses incident on the BNA-S crystal, the emitted THz pulse energy was ~10 nJ, corresponding to a peak electric field of ~40kV/cm.

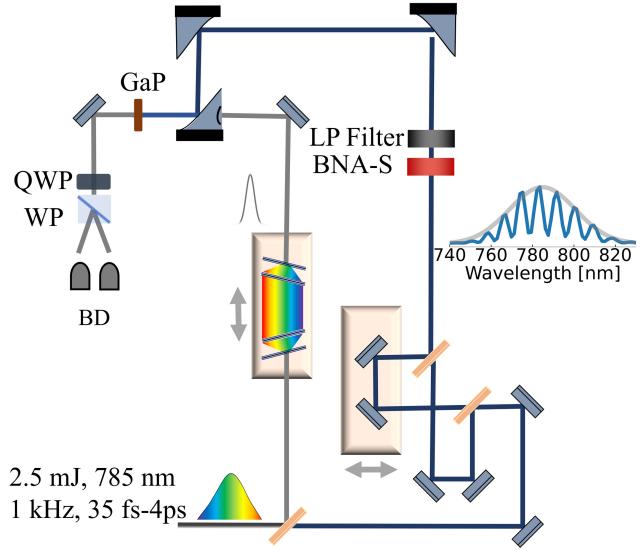

**Figure 2:** Schematic of the setup. The output of a Ti:sapphire laser system is linearly chirped by detuning the compressor grating separation away from its optimal value. A beamsplitter, transmits 1% of the laser output, which is fully compressed back to transform-limit configuration using a transmission grating compressor. The remaining 99% is directed to a variable delay Mach-Zehnder Interferometer (MZI), in which pulses are split and delayed, creating a temporally modulated intensity envelope with a beat frequency $\Omega$. The narrowband THz pulses emitted by the BNA-S crystal are expanded with an OAP reflective telescope and focused onto a 100-um thick GaP (110) crystal where they are sampled. A quarter-wave plate (QWP), Wollaston prism (WP) and balanced detector (BD) are used to detect the THz induced Pockels effect in GaP. The residual pump after BNA-S crystal is blocked by a stack of low pass filters (LP). The inset shows the intensity spectrum recorded at the MZI output with a fiber-coupled spectrometer.

Figures 3(a) and 3(b) show EOS time-domain waveforms and corresponding spectra for three representative optical pulse durations: 35 fs, 1.2 ps and 4 ps. For each case, the interferometer delay $\tau$ was adjusted to maintain the center frequency approximately constant at ~1 THz. Increasing the GDD introduced in the optical pulses – and thus their duration – progressively narrows the THz bandwidth. This behavior directly reflects the principle illustrated in Fig. 1: as the the linear chirp parameter increases, fewer optical frequency components overlap in time, resulting in a correspondingly narrower THz spectrum.

The tunability of the generated THz pulses was investigated for constant duration of the optical pulses. Fig. 3(c, d) show the THz waveforms and corresponding spectra acquired for a fixed duration of the optical pulses of 4 ps, varying only the interferometer delay $\tau$. While the overall duration of the THz transient remains roughly constant, its central frequency (Fig. 3(d)) increases for increasing delays. Fig. 3(e) showcases the continuous tunability of the source in the range between 0.25 and 2.1 THz. Throughout the range, the THz bandwidth also increases from ~ 0.2 to 0.4 THz, likely due to a reduction of the duration of the generated THz radiation arising from the $\sqrt{I_1(t)I_2(t+\tau)}$ term of Eq. (1). For comparison, the dashed line shows the THz spectrum generated by the transform-limited 35 fs long optical pulses in the case of intra-pulse DFG (Fig.1(a,b)). The temporal shaping of the pump pulses is instrumental in mitigating nonlinear absorption in BNA, which is key both for conversion efficiency and long-term crystal stability. To elucidate this effect, fluence-dependent transmission measurements were performed using fully compressed (~35 fs) pulses and chirped pulses with durations of 1.2 ps, 2.2 ps, and 4 ps.

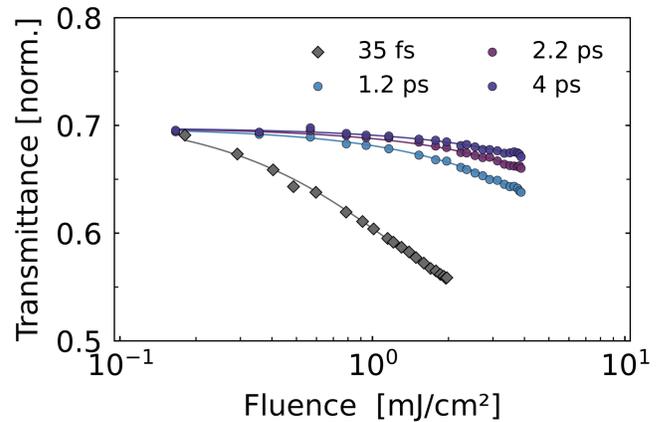

**Figure 4:** Fluence-dependent transmission of the BNA-S crystal using both compressed pulses (~35 fs) and chirped pulses with durations of 1.2 ps, 2.2 ps, and 4 ps. The 35 fs data were fitted using a saturated two-photon absorption model, while a simple two-photon absorption model was applied to the chirped-pulse data [30].

As shown in Fig. 4, the nonlinear absorption response differs markedly between the two cases. Fully compressed pulses exhibit pronounced transmission loss with increasing fluence, consistent with multiphoton absorption (MPA) [28-30], whereas chirped pulses sustain substantially higher transmission under equivalent excitation conditions, yielding higher conversion efficiency and extending the crystal lifetime.

In summary, we described a source of narrowband THz radiation based on difference-frequency generation in the organic crystal BNA-S, pumped directly by the fundamental output of a Ti:sapphire amplifier. The use of inter-pulse DFG between delayed chirped pulses enables continuous tunability of both center frequency in the range between 0.25-2.2 THz and spectral bandwidth. In addition, exciting BNA-S with chirped pulses lowers drastically non-linear absorption in the crystal improving its longevity and stability.

**Data availability.** Data underlying the results presented in this paper are not publicly available but may be obtained from the authors upon reasonable request.

**Acknowledgment.** D.P. acknowledges support by the Max Planck Graduate Center for Quantum Materials (MPGC-QM). We are also grateful to Boris Fiedler, Birger Höhling and Toru Matsuyama for their support in the fabrication of the electronic devices used in the measurement setup.

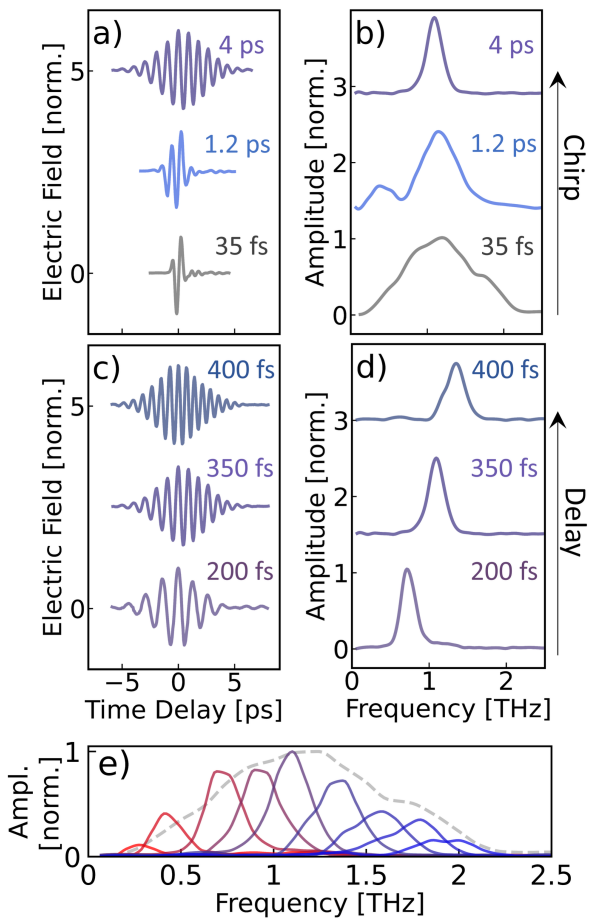

**Figure 3:** **(a)** EOS traces obtained for three representative optical pulse durations (35 fs, 1.2 ps and 4 ps). The interferometer delay $\tau$ was adjusted to keep the THz center frequency near 1 THz. **(b)** Corresponding Fourier transform spectra of the waveforms in (a). **(c)** EOS traces measured at a fixed optical pulse duration of 4 ps for three representative delay values $\tau$ = 0.2 ps, 0.35 ps, and 0.4 ps. **(d)** Fourier transform spectra corresponding to the THz waveforms shown in (c). **(e)** Collection of THz spectra measured varying the interferometer delay $\tau$ showing continuous frequency tunability in the range from 0.25 to 2.1 THz for 4 ps chirped pulses. The dashed line is the broadband THz spectrum generated with fully compressed optical pulses with 35 fs duration.

**References**

1. D. Nicoletti, A. Cavalleri, "Nonlinear light–matter interaction at terahertz frequencies, "Adv. Opt. Photon. **8**, 401-464 (2016)
2. T. Kampfrath, A. Sell, G. Klatt, *et al*., "Coherent terahertz control of antiferromagnetic spin waves, " Nature Photon 5, 31–34 (2011).
3. Z. Zhang, F.Y. Gao, J.B. Curtis, *et al.,* "Terahertz field-induced nonlinear coupling of two magnon modes in an antiferromagnet," Nat. Phys. **20**, 801–806 (2024).
4. T. Kubacka, J.A. Johnson, M.C. Hoffmann, *et al*., "Large-Amplitude Spin Dynamics Driven by a THz Pulse in Resonance with an Electromagnon.," Science **343**, 1333–1336 (2014).
5. Y. Laplace, A. Cavalleri, "Josephson plasmonics in layered superconductors, " Advances in Physics: X, 1(3), 387–411 (2016).

6. A. Dienst, M. Hoffmann, D. Fausti, *et al.,* "Bi-directional ultrafast electric-field gating of interlayer charge transport in a cuprate superconductor, " Nature Photon **5**, 485–488 (2011).
7. E.J. Sie, C. M. Nyby, C.D. Pemmaraju, *et al.*, "An ultrafast symmetry switch in a Weyl semimetal, " Nature **565**, 61–66 (2019).
8. F. Blanchard, L. Razzari, H. C. Bandulet, *et al*., "Generation of 1.5 μJ single-cycle terahertz pulses by optical rectification from a large aperture ZnTe crystal, " Opt. Express 15(20), 13212–13220 (2007).
9. M. C. Hoffmann, K.-L. Yeh, J. Hebling, et al., "Efficient terahertz generation by optical rectification at 1035 nm," Opt. Express 15, 11706–11713 (2007).
10. K.Aioki, J. Savolainen, M. Havenith, "Broadband terahertz pulse generation by optical rectification in GaP crystals, " Appl. Phys. Lett. 110, 201103 (2017).
11. K.-L. Yeh, M. C. Hoffmann, J. Hebling, *et al*., "Generation of 10 μJ ultrashort terahertz pulses by optical rectification, " Appl. Phys. Lett. 90(17), 171121 (2007).
12. J. Hebling, K.-L. Yeh, M. C. Hoffmann, *et al*., "Generation of high-power terahertz pulses by tilted-pulse-front excitation and their application possibilities," J. Opt. Soc. Am. B 25, B6-B19 (2008).
13. S.-W. Huang, E. Granados, W. Ronny, *et al*., "High conversion efficiency, high energy terahertz pulses by optical rectification in cryogenically cooled lithium niobate, " Opt. Lett. 38, 796–798 (2013).
14. B. Zhang, Z. Ma, J, Ma, *et al*., "1.4-mJ high energy terahertz radiation from lithium niobates, " Laser & Photonics Reviews, 15, 2000295 (2021).
15. B. Monoszlai, C. Vicario, M. Jazbinsek, *et al.,* "High-energy terahertz pulses from organic crystals: DAST and DSTMS pumped at Ti:sapphire wavelength," Opt. Lett. **38**, 5106-5109 (2013).
16. C. P. Hauri, C. Ruchert, C. Vicario, *et al.*, "Strong-field single-cycle THz pulses generated in an organic crystal," Appl. Phys. Lett. 99(16), 161116 (2011).
17. C. Vicario, A.V. Ovchinnikov, S.I. Ashitkov, *et al*., "Generation of 0.9-mJ THz pulses in DSTMS pumped by a Cr:Mg$_2$SiO$_4$ laser," Opt. Lett. 39(23), 6632-5 (2014).
18. C. Vicario, M. Jazbinsek, A. V. Ovchinnikov, *et al*., "High efficiency THz generation in DSTMS, DAST and OH1 pumped by Cr:forsterite laser," Opt. Express 23, 4573-4580 (2015).
19. M. Shalaby, C. Vicario, K.Thirupugalmani, *et al.,* "Intense THz source based on BNA organic crystal pumped at Ti:sapphire wavelength," Opt. Lett. 41, 1777-1780 (2016).
20. H. Zhao, Y. Tan, T. Wu, et al., "Efficient broadband terahertz generation from organic crystal BNA using near infrared pump, " Appl. Phys. Lett. 14 (24): 241101 (2019).
21. F. Roeder, M. Shalaby, B. Beleites, et al., "THz generation by optical rectification of intense near-infrared pulses in organic crystal BNA," Opt. Express 28, 36274-36285 (2020).
22. I. C. Tangen, G. A. Valdivia-Berroeta, L. K. Heki, et al., "Comprehensive characterization of terahertz generation with the organic crystal BNA," J. Opt. Soc. Am. B 38, 2780-2785 (2021)
23. J. R. Danielson, A. D. Jameson, J. L. Tomaino, *et al*., "Intense narrow band terahertz generation via type-II difference-frequency generation in ZnTe using chirped optical pulses, " J. Appl. Phys*,* 104 (3), 033111 (2008).
24. Z. Chen, X. Zhou, C. A. Werley, et al., "Generation of high power tunable multicycle teraherz pulses, "Appl. Phys. Lett., 99 (7), 071102 (2011).
25. C. Vicario, A. Trisorio, S. Allenspach, et al., "Narrow-band and tunable intense terahertz pulses for mode-selective coherent phonon excitation, " Appl. Phys. Lett*.,* 117 (10), 101101 (2020).
26. A. V. Ovchinnikov, O. V. Chefonov, M. B. Agranat, et al., "Generation of strong-field spectrally tunable terahertz pulses," Opt. Express 28, 33921-33936 (2020).
27. J. Lu, H. Y. Hwang, X. Li, et al., "Tunable multi-cycle THz generation in organic crystal HMQ-TMS," Opt. Express 23, 22723-22729 (2015).
28. R. Schroeder and B. Ullrich, "Absorption and subsequent emission saturation of two-photon excited materials: theory and experiment, " Opt. Lett. 27(15), 1285 (2002).
29. G. S. He, L.-S. Tan, Q. Zheng, et al., "Multiphoton Absorbing Materials: Molecular Designs, Characterizations, and Applications," Chemical Reviews 108 (4), 1245-1330 (2008)
30. A. Gupta, T. Zhang, V. Hanyecz, et al., "Two-photon absorption and its saturation in organic terahertz-generator crystals," Opt. Mater. Express 15, 2056-2065 (2025).